\documentclass[aps,prl,twocolumn,superscriptaddress,floatfix,nofootinbib]{revtex4-2}

\usepackage{amsmath,amssymb,bm}
\usepackage{booktabs}
\usepackage{array}
\usepackage{tikz}
\usepackage{microtype}
\usepackage[colorlinks=true,citecolor=blue,linkcolor=blue,urlcolor=blue]{hyperref}

\newcommand{\ii}{\mathrm{i}}
\newcommand{\dd}{\mathrm{d}}
\newcommand{\tr}{\operatorname{tr}}
\newcommand{\SU}{\mathrm{SU}}
\newcommand{\U}{\mathrm{U}}
\newcommand{\Z}{\mathbb Z}
\newcommand{\cL}{\mathcal L}
\newcommand{\cO}{\mathcal O}
\newcommand{\hc}{\mathrm{h.c.}}
\newcommand{\vev}[1]{\left\langle #1\right\rangle}
\newcommand{\gauge}{\mathrm g}
\newcommand{\val}{\mathrm v}

\begin{document}

\title{Color superconductors and holon metals from doping a Fractional Chern insulator}

\author{Ya-Hui Zhang}
\affiliation{Department of Physics and Astronomy, Johns Hopkins University, Baltimore, Maryland 21218, USA}

\date{\today}

\begin{abstract}
We develop a unified framework for metallic and superconducting phases
obtained by doping a fractional Chern insulator (FCI) with $C=1/3$.
Starting from the parton construction
$c(\mathbf r)=f_1(\mathbf r)f_2(\mathbf r)f_3(\mathbf r)$, the low-energy
theory has $SU(3)_{\mathrm{gauge}}\times SU(3)_{\mathrm{valley}}$ symmetry
and nine Fermi pockets formed by charge-$-e/3$ holons $\psi_{ab}$, where
$a$ and $b$ label color and valley.  Viewing the holons as quarks connects
this problem to color superconductivity in high-energy physics.
Color-antisymmetric pairing produces a class of charge-$2e$
superconductors with angular momentum $L=3n$ and chiral central charge
$c_-=m/2$, where $m$ is odd.  Thus a gas of charge-$e/3$ anyons can enter a
superconducting phase directly without binding.  Particle--hole color--valley Higgs fields
instead produce two $Z_3$ orthogonal metals with one or three pockets,
transforming respectively as a singlet or triplet of
$SU(3)_{\mathrm{valley}}$.  A $U(1)^2$ holon metal with three identical
pockets can preserve the triangular-lattice space group while reducing the
emergent valley symmetry down to $S_3$.  Its pairing instabilities include a gapped charge $2e$ $f-if$ superconductor and a gapless
charge-$2e$ orthogonal superconductor with $\langle cc\rangle=0$ and a
Bogoliubov Fermi surface at $\Gamma$.  Finally, we discuss the possibility of a
chemical-potential-tuned transition from the FCI to superconductivity and
argue that all nine fermions may be required if the transition preserves
the full emergent $SU(3)_v$ symmetry.
\end{abstract}

\maketitle

\emph{Introduction.---}
Fractional Chern insulators (FCIs) combine fractional quantum Hall
topological order with microscopic lattice symmetry.  The possibility that a
finite-density anyon fluid becomes superconducting---usually called anyon
superconductivity---was developed in early work
\cite{Laughlin1988,LeeFisher1989,Fetter1989,Chen1989,WenZee1990,TangWen2013}
and has recently attracted renewed attention
\cite{ShiSenthil2025,Divic2025,Kim2025,Zhang2025Holon,Pichler2026,
ShiZhangSenthil2025,Nosov2026,ShiSenthilDisorder2025,Han2026,
Nakajima2025,Kuhlenkamp2025,ShiSenthil2026NonAbelian,Lotric2026,
Fan2026,Wang2026Topological,ShiSenthil2026Fluid,Senthil2026Fractionalized,
HanVishwanathKhalaf2026}.
This revival is motivated in part by the observation of superconductivity
near fractional quantum anomalous Hall states in twisted MoTe$_2$
\cite{Xu2025Experiment}.  Independent numerical studies have also found
chiral $f$-wave superconductivity near an FCI
\cite{Guerci2025,WangZaletel2025}, although it is still unclear whether
anyons are relevant to the pairing mechanism.  Here we consider doping an FCI with
$\sigma_{xy}=\frac{1}{3}\frac{e^2}{h}$ from $n=1/3$ to $n=1/3-x$.
A conventional Fermi liquid may emerge for $x>x_c$, whereas the underdoped
region $x<x_c$ may be better described as a gas of anyons.  The value of
$x_c$, and whether the experimental superconductor emerges from a
conventional Fermi liquid or an anyon gas, remain open questions.  A complete
phase diagram must ultimately capture the crossover between these regimes,
which is beyond the scope of this work.

We assume that the only dynamical degrees of freedom are the doped holes,
with number density $n_h=x$ (equivalently, electron-density change
$\delta n=-x$), while the original
$N_e=\frac{1}{3}N_s$ electrons remain nearly inert.  Strictly speaking, this
assumption is controlled only when the number of doped anyons is much smaller
than the number of sites $N_s$, namely in the $x\to0$ limit.  We nevertheless
focus on the doped anyons at finite but small $x$.  A common approach
represents an anyon as a bosonic or fermionic holon of charge $-e/3$ coupled
to a Chern--Simons gauge field.  At finite $x$, these holons experience an
effective magnetic flux and may form quantum Hall states.  For a doped
$\nu=1/2$ bosonic Laughlin state, the holons can form a gapped integer quantum Hall state, and
the resulting physical phase is a superfluid.  In a doped $n=1/3$ FCI,
however, the holons generically occur at an effective filling for which no
simple gapped quantum Hall state exists.  They may instead form a composite
Fermi liquid, yielding a metallic phase.  Intuitively, three such anyons can
combine into a fermion, so a metal is a natural possibility; a
superconductor may then be reached through a pairing instability.  Different
composite-fermion or composite-boson constructions can nevertheless produce
different physical states, and their relation is not always transparent.
Moreover, microscopic space-group and emergent symmetries can be difficult to
track in these descriptions.  A unified framework that makes these symmetries
explicit is therefore desirable.

We show that the parton approach provides such a framework and also yields
phases not readily visible in the earlier constructions.  We begin with
$c(\mathbf r)=f_1(\mathbf r)f_2(\mathbf r)f_3(\mathbf r)$ and an
$SU(3)_g$ gauge symmetry.  At $x=0$, each $f_a$ occupies a $C=1$ Chern
insulator, producing the familiar Laughlin wavefunction.  At finite $x$, each
color can host three hole pockets at valleys related by projective
translations.  We denote the low-energy hole operators by $\psi_{ab}$, where
$a=1,2,3$ is the color index and $b=1,2,3$ is the valley index.  Each
$\psi_{ab}$ can be viewed as a charge-$-e/3$ holon.  Starting from a holon gas
with $SU(3)_g$ gauge symmetry and a global $SU(3)_v$ symmetry rotating the
valleys, we obtain distinct superconducting and metallic phases by adding
particle--hole and particle--particle holon bilinears.  The same construction
also supplies a flexible family of unified variational wavefunctions for
future numerical studies.

First, an anyon superconductor can arise through a mechanism closely analogous
to color superconductivity in high-energy physics
\cite{Alford1999,Alford2008}.  The $SU(3)$ gauge field mediates pairing in the
color-antisymmetric channel.  Because the present problem is spinless, an
odd-angular-momentum holon pair may be antisymmetric in both color and valley.
The pair condensate locks $SU(3)_g$ to $SU(3)_v$, completely Higgses the gauge
symmetry, and leaves a new global $SU(3)_v$.  The result is a charge-$2e$
superconductor.  In the convention used throughout this paper, holon
$p+ip$ contributes $c_-^{\rm pocket}=-1/2$ and corresponds to
physical-electron $f-if$, whereas holon $p-ip$ contributes
$c_-^{\rm pocket}=+1/2$ and corresponds to physical-electron $f+if$.
For example, for holon $p+ip$ and electron $f-if$, the
chiral central charge is $c_-=3-N_\psi/2$, where $N_\psi$ is the number of
weakly paired holon pockets.  Pairing all nine pockets gives
$c_-=-3/2$, distinct from the BCS expectation for an $f-if$
superconductor.  Thus the usual relation between $c_-$ and the angular
momentum of the physical pair need not hold in an anyon superconductor.
Symmetry also allows $N_\psi$ to vary.  Under the locked $SU(3)_v$, the nine
holons decompose as $1\oplus8$.  We denote the singlet by $\psi_0$ and the
octet by $\psi_i$, $i=1,\ldots,8$.  Symmetry permits $\psi_0$ to lie below the
octet, giving $N_\psi=1$.  If $SU(3)_v$ is reduced to the triangular-lattice
space group, the octet further splits as $8=2\oplus6$, allowing
$N_\psi=3$ for weakly paired $\psi_0,\psi_1,\psi_2$.  The $p+ip$ ansatz
then gives $c_-=3/2$: the three holon pockets contribute $-3/2$, while the
filled-band background contributes $+3$.  This matches the BCS value for the 
$f-if$ pairing, and the anyon superocnductor may cross over to a BCS state.  The
color-superconductor framework therefore produces a flexible family of
charge-$2e$ superconductors that can be studied by variational Monte Carlo.

Intermediate metallic phases formed by the charge-$-e/3$ holons are also
possible.  These holon metals are distinguished by their gauge structure and
global symmetry.  Starting from the $SU(3)_g\times SU(3)_v$ holon metal, we
obtain descendants by adding particle--hole Higgs terms.  Preserving
$SU(3)_v$ permits two $Z_3$ holon metals with one or three pockets,
corresponding to ordinary or conjugate locking of $SU(3)_g$ to $SU(3)_v$.
These two  states should be equivalent to the $Z_3$ orthogonal metals
recently discussed in
Refs.~\cite{ShiSenthil2026Fluid,Senthil2026Fractionalized,HanVishwanathKhalaf2026}.
If $SU(3)_v$ is reduced to the triangular-lattice space group, a $U(1)^2$
holon metal with three translation-related pockets becomes possible.
Closely related $U(1)^2$ holon metals were constructed in
Refs.~\cite{Zhang2025Holon,Fan2026}, but with different projective
translations.  Those realizations necessarily break $C_3$, whereas the
construction developed here preserves the full triangular-lattice space
group.  We also discuss pairing instabilities of this phase.
Translation-preserving $p+ip$ pairing gives the same superconductor as the
color-superconductor construction with $N_\psi=3$, whereas $s$-wave pairing produces a
gapless charge-$2e$ superconductor with a remaining Bogoliubov Fermi surface.

To assess which metallic parent is more natural, we briefly discuss the
chemical-potential-tuned transition from the FCI to a superconductor.  The
$Z_3$ metals cannot be reached from the FCI directly and an intermediate phase is required.  By contrast, a
continuous FCI--SC transition may occur when the superconductor descends from
the $SU(3)$ or $U(1)^2$ holon metal.  The $U(1)^2$ holon metal, however,
necessarily breaks the emergent $SU(3)_v$ symmetry and cannot preserve the
full lattice symmetry on the square lattice.  The full nine-pocket $SU(3)$
theory may therefore provide the natural starting point, with the other
phases reached as its descendants.

\textit{Parton construction.---} We consider density $n=\frac{1}{3}-x$
in a $C=1$ band.  Our starting point is
\begin{equation}
 c(\bm r)=f_1(\bm r)f_2(\bm r)f_3(\bm r),
\label{eq:parton}
\end{equation}
where $f_a$ carries electric charge $e/3$.  A local color rotation
$f\mapsto Uf$, $U\in\SU(3)_{\gauge}$, leaves the electron invariant.  The
physical on-site Hilbert space therefore contains only an empty site and a
color singlet with all three colors occupied.  We retain both
particle--hole and particle--particle variational channels in the quadratic
parton mean-field Hamiltonian
\begin{align}
 H_{\rm var}={}&H_{\rm Hof}-\mu_fN_f
 +\sum_{\bm k}\Psi_{\bm k}^\dagger
 {\cal M}_{\rm ph}(\bm k)\Psi_{\bm k}\nonumber\\[-2pt]
 &+\frac12\sum_{\bm k}\left[
 \Psi_{\bm k}^{T}\Delta_{\rm pp}(\bm k)\Psi_{-\bm k}+\hc\right].
\label{eq:unified-mean-field}
\end{align}
Here $H_{\rm Hof}$ is a $2\pi/3$-flux per unit cell Hofstadter
ansatz to generate a $C=1$ band for each color.  ${\cal M}_{\rm ph}$ is a particle--hole Higgs field, and
$\Delta_{\rm pp}$ is a particle--particle Higgs field.  A variational
electron wavefunction at $n=\frac{1}{3}-x$ is then simply
$\Phi_{\rm G}^{f_1f_2f_3}(\bm r_1,\ldots,\bm r_{N_e};
{\cal M}_{\rm ph},\Delta_{\rm pp})$, the fixed-number coordinate amplitude
of the Gaussian ground state evaluated on configurations with one
$f_1,f_2,f_3$ triplet at each $\bm r_i$, where $N_e=(1/3-x)N_s$.  At
$x=0$, it reduces to the familiar Laughlin wavefunction when
${\cal M}_{\rm ph}=\Delta_{\rm pp}=0$ and each $f_a$ fills a $C=1$ band.

At $x=0$, magnetic translations act projectively on each color, producing
three hole valleys.  We denote the corresponding hole
annihilators by $\psi_{ab}$, with color $a$ and valley $b$, and organize the
nine fields as the matrix $\Psi=(\psi_{ab})$.  A low-energy effective theory is
\begin{align}
 \cL_0={}&\tr\!\left\{
 \Psi^\dagger\left[\ii D_t+\frac{\bm D^2}{2m_*}+\mu\right]\Psi\right\}
 \nonumber\\
 &+\frac{1}{4\pi}\tr\!\left(a\dd a-\frac{2\ii}{3}a^3\right)
 +\frac{1}{12\pi}A\dd A,
\label{eq:SU3-critical-theory}
 \\
 D_\mu\Psi={}&(\partial_\mu-\ii a_\mu+\ii A_\mu/3)\Psi .
\label{eq:SU3-covariant-derivative}
\end{align}
The trace in the first term is over both indices, whereas that in the
Chern--Simons term is over color.  Here $a_\mu$ is the $SU(3)$ gauge field
and $A_\mu$ is the electromagnetic probe field.  The action is written in
real time.  We use
$a\dd b\equiv\epsilon_{\mu\nu\sigma}a_\mu\partial_\nu b_\sigma$, with
$\epsilon_{012}=+1$.  The FCI is the vacuum
of hole excitations ($\mu<0$), whereas $\mu>0$ creates nine small pockets
with total density $n_\psi=3x$.  Equation~\eqref{eq:SU3-critical-theory}
is the starting point for adding particle--hole and particle--particle holon
bilinears.

The continuum theory has an $SU(3)_g$ gauge symmetry and an emergent global
$\SU(3)_{\val}$ rotating the valleys.  Their action is
$\Psi\mapsto U\Psi V^\dagger$, where $U\in\SU(3)_{\gauge}$ and
$V\in\SU(3)_{\val}$.  The microscopic triangular-lattice space group is
embedded in $SU(3)_v$.  For example, the translation matrices obey
$V_{T_1}V_{T_2}=\omega V_{T_2}V_{T_1}$,
$\omega=e^{2\pi\ii/3}$.

\textit{Color superconductor.---} We consider a close analog of color
superconductivity in a quark gas.  The $SU(3)_g$ gauge field mediates an
attraction and favors the pairing
\begin{equation}
 \vev{\psi_{ab}(\bm k)\psi_{a'b'}(-\bm k)}
 =\Delta\varphi_\ell(\bm k)
 \sum_{A=1}^{3}\epsilon_{Aaa'}\epsilon_{Abb'} .
\label{eq:CFL-pair}
\end{equation}
Because we do not have spin, odd orbital parity is compatible with an
antisymmetric tensor in both color and valley.  Define
$P_{AB}=\epsilon_{Aaa'}\epsilon_{Bbb'}\psi_{ab}\psi_{a'b'}$.  Then
$P\mapsto U^*PV^T$ under $\SU(3)_{\gauge}\times\SU(3)_{\val}$.  A
full-rank condensate $\Phi_{AB}=\langle P_{AB}\rangle=\Phi\delta_{AB}$
completely Higgses color and leaves a locked global group generated by a
valley operation accompanied by $U=V$.  On the fermions it acts as
$\Psi\mapsto V\Psi V^\dagger$.

The fractional pair field $\Phi$ is gauge charged and is not itself an
electron order parameter.  The corresponding gauge-invariant charge-$2e$
order parameter is $\cO_{2e}=\frac1{3!}\epsilon^{ABC}\epsilon^{A'B'C'}
 \Phi^*_{AA'}\Phi^*_{BB'}\Phi^*_{CC'}
 =(\det\Phi)^\dagger\sim cc$.  Consequently $\cO_{2e}$ has charge $2e$
and angular momentum $L_e=-3\ell$,
where $\ell$ is the angular momentum of the holon pairing.  The Chern--Simons
term selects $p+\ii p$ over $p-\ii p$ for the holons and produces
$f-\ii f$ pairing of the physical electrons.  The chiral central charge is
$c_-=3-N_\psi\ell/2$, where $3$ is from the $C=3$ of the filled Chern band at $x=0$ and $N_\psi$ is the number of holon pockets in the weak pairing phase. For the $p+\ii p$ pairing, we have $c_-=-3/2$ when all nine pockets are paired.

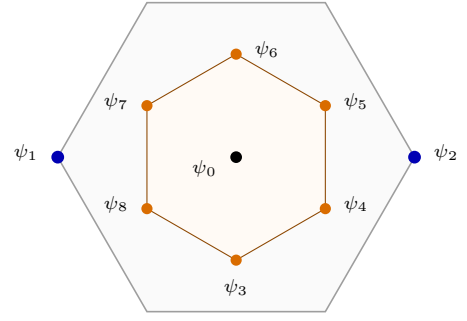
\begin{figure}[t]
\centering
\begin{tikzpicture}[
  x=1.18cm,y=1.18cm,
  every node/.style={font=\scriptsize}
]
  \fill[black!2]
    (2,0)--(1,{sqrt(3)})--(-1,{sqrt(3)})--(-2,0)--
    (-1,{-sqrt(3)})--(1,{-sqrt(3)})--cycle;
  \draw[black!38,semithick]
    (2,0)--(1,{sqrt(3)})--(-1,{sqrt(3)})--(-2,0)--
    (-1,{-sqrt(3)})--(1,{-sqrt(3)})--cycle;
  \fill[orange!4]
    (0,{-2/sqrt(3)})--(1,{-1/sqrt(3)})--(1,{1/sqrt(3)})--
    (0,{2/sqrt(3)})--(-1,{1/sqrt(3)})--(-1,{-1/sqrt(3)})--cycle;
  \draw[orange!55!black,thin]
    (0,{-2/sqrt(3)})--(1,{-1/sqrt(3)})--(1,{1/sqrt(3)})--
    (0,{2/sqrt(3)})--(-1,{1/sqrt(3)})--(-1,{-1/sqrt(3)})--cycle;

  \fill (0,0) circle (2.2pt);
  \node[anchor=east] at (-0.12,-0.13) {$\psi_0$};

  \fill[blue!70!black] (-2,0) circle (2.4pt);
  \node[anchor=east] at (-2.12,0.05) {$\psi_1$};

  \fill[blue!70!black] (2,0) circle (2.4pt);
  \node[anchor=west] at (2.12,0.05) {$\psi_2$};

  \fill[orange!85!black] (0,{-2/sqrt(3)}) circle (2.1pt);
  \fill[orange!85!black] (1,{-1/sqrt(3)}) circle (2.1pt);
  \fill[orange!85!black] (1,{1/sqrt(3)}) circle (2.1pt);
  \fill[orange!85!black] (0,{2/sqrt(3)}) circle (2.1pt);
  \fill[orange!85!black] (-1,{1/sqrt(3)}) circle (2.1pt);
  \fill[orange!85!black] (-1,{-1/sqrt(3)}) circle (2.1pt);

  \node[anchor=north] at (0,-1.27) {$\psi_3$};

  \node[anchor=west] at (1.11,{-1/sqrt(3)+0.05}) {$\psi_4$};

  \node[anchor=west] at (1.11,{1/sqrt(3)+0.05}) {$\psi_5$};

  \node[anchor=west] at (0.11,{2/sqrt(3)+0.05}) {$\psi_6$};

  \node[anchor=east] at (-1.11,{1/sqrt(3)+0.05}) {$\psi_7$};

  \node[anchor=east] at (-1.11,{-1/sqrt(3)+0.05}) {$\psi_8$};
\end{tikzpicture}
\caption{\label{fig:nine-field-momenta}
Momentum assignment of the nine fields in
Eq.~\eqref{eq:explicit-one-eight}.  The black center is the singlet under the $SU(3)_v$ symmetry after the locking $U=V$, the blue pair is the
$C_6$ orbit $(\psi_1,\psi_2)$, and the orange inner hexagon is the six-field
orbit $(\psi_3,\ldots,\psi_8)$.  }
\end{figure}

Symmetry allows $N_\psi$ to differ from $9$.  Under the new $SU(3)_v$, the
nine holon fields decompose into the $1\oplus8$ representation.  With
$\omega=e^{2\pi\ii/3}$, we redefine the nine holons as
\begin{equation}
\begin{alignedat}{2}
 \psi_0&=\frac{\psi_{11}+\psi_{22}+\psi_{33}}{\sqrt3},\\
 \psi_1&=\frac{\omega\psi_{12}+\psi_{23}+\omega^2\psi_{31}}{\sqrt3},&\quad
 \psi_2&=\frac{\omega^2\psi_{21}+\psi_{32}+\omega\psi_{13}}{\sqrt3},\\
 \psi_3&=\frac{\psi_{12}+\psi_{23}+\psi_{31}}{\sqrt3},&\quad
 \psi_4&=\frac{\psi_{11}+\omega^2\psi_{22}+\omega\psi_{33}}{\sqrt3},\\
 \psi_5&=\frac{\omega\psi_{21}+\psi_{32}+\omega^2\psi_{13}}{\sqrt3},&\quad
 \psi_6&=\frac{\psi_{21}+\psi_{32}+\psi_{13}}{\sqrt3},\\
 \psi_7&=\frac{\psi_{11}+\omega\psi_{22}+\omega^2\psi_{33}}{\sqrt3},&\quad
 \psi_8&=\frac{\omega^2\psi_{12}+\psi_{23}+\omega\psi_{31}}{\sqrt3}.
\end{alignedat}
\label{eq:explicit-one-eight}
\end{equation}
Here $\psi_0$ is the $SU(3)_v$ singlet.  In this locked basis, the
space-group operations $T_1,T_2,C_6$ can be diagonalized simultaneously,
and the nine fields can be assigned momenta in the Brillouin zone (BZ), as
shown in Fig.~\ref{fig:nine-field-momenta}.  The singlet $\psi_0$ lies at
$\Gamma$, while $\psi_1$ and $\psi_2$ lie at $K$ and $K'$.  The remaining
six fermions lie at the six momenta $\mathbf G/3$.  Space-group operations
therefore act in the usual way in this basis.

The color-antisymmetric interaction gaps all nine pockets, with
\begin{equation}
 \boxed{\Delta_0=2\Delta,\qquad
 \Delta_{12}=\Delta_{36}=\Delta_{47}=\Delta_{58}=-\Delta}.
\label{eq:gap-relations}
\end{equation}
Here $\Delta_0$ is the intrapocket pairing of $\psi_0$, and $\Delta_{ij}$
is the pairing between the $\psi_i$ and $\psi_j$ pockets.

The $SU(3)_v$ symmetry also permits the singlet to be lowered relative to
the octet through, for example,
$\delta H=-\delta\psi_0^\dagger\psi_0+
\delta\sum_{i=1}^8\psi_i^\dagger\psi_i$, yielding $N_\psi=1$.  If
$SU(3)_v$ is reduced to the space group, $p+ip$ pairing within $\psi_0$ and
between $\psi_1,\psi_2$ gives $N_\psi=3$ and $c_-=3/2$, matching the BCS
value for an $f-if$ superconductor.

\textit{Holon metals.---} Above the superconducting $T_c$, a
non-Fermi-liquid metal may be formed by charge-$-e/3$ holons rather than
charge-$e$ electrons.  The first example is the $\SU(3)_{\gauge}$
holon metal with nine pockets and full $\SU(3)_{\val}$ symmetry.  We can
have other types of holon metals with a smaller gauge symmetry by adding a
particle--hole Higgs field.  If we preserve the full $\SU(3)_{\val}$ symmetry, the gauge symmetry
must be Higgsed down to $\Z_3$.  We obtain two distinct $\Z_3$ holon
metals, also called orthogonal metals in recent
works~\cite{ShiSenthil2026Fluid,Senthil2026Fractionalized,HanVishwanathKhalaf2026}.
The two possibilities are
\begin{itemize}
\item \emph{$\Z_3$ holon metal with one pocket.---}
We consider the ordinary locking $U=V$.  Choose generators normalized by
$\tr(T^A T^B)=\delta^{AB}/2$ and use the same adjoint label $A$ for color
and valley.  In matrix notation the Higgs term is
\begin{equation}
 \delta H_{\rm ord}
 =-\lambda\sum_{A=1}^{8}
 \tr_{\val}\!\left(\Psi^\dagger T^A_{\gauge}\Psi T^A_{\val}\right)
 =-\frac{4\lambda}{3}n_0+\frac{\lambda}{6}n_8 .
\label{eq:ordinary-Z3-Higgs}
\end{equation}
It is invariant under $\Psi\mapsto U\Psi V^\dagger$ when
$\mathrm{Ad}_U=\mathrm{Ad}_V$, or $U=zV$ with $z\in\Z_3$.  Thus its
continuous unbroken group is the ordinary diagonal $U=V$, while the
gauge-only invariance is the center $\Z_3$.  Here
$n_0=\psi_0^\dagger\psi_0$ and
$n_8=\sum_{i=1}^8\psi_i^\dagger\psi_i$.
It splits
${\bm3}_{\gauge}\otimes\overline{\bm3}_{\val}={\bm1}\oplus{\bm8}$;
for $\lambda>0$ this leaves the pocket
$\psi_0=\tr\Psi/\sqrt3$, coupled to a remaining $\Z_3$ gauge field.
The pocket is at the $\Gamma$ point and is a singlet under the new
$\SU(3)_{\val}$ symmetry.  Pairing from this metal leads to the same
superconductor as the color-superconductor ansatz with $N_\psi=1$.  

\item \emph{$\Z_3$ holon metal with three pockets.---}
We consider the conjugate locking $U=V^*$.  It differs from the ordinary
lock by transposing the valley generator in the matrix contraction:
\begin{equation}
 \delta H_*
 =+\lambda\sum_{A=1}^{8}
 \tr_{\val}\!\left[\Psi^\dagger T^A_{\gauge}\Psi
 (T^A_{\val})^T\right].
\label{eq:conjugate-Z3-Higgs}
\end{equation}
It is invariant when
$\mathrm{Ad}_U=\mathrm{Ad}_{V^*}$, or $U=zV^*$, and hence realizes the
conjugate diagonal lock.  
It splits
$\overline{\bm3}\otimes\overline{\bm3}={\bm3}_{a}\oplus
\overline{\bm6}_{s}$ under the new $\SU(3)_{\val}$ symmetry, for which
$\Psi\mapsto V^*\Psi V^\dagger$.  For $\lambda>0$ the three pockets are
\begin{equation}
 \eta_A=\frac1{\sqrt2}\epsilon_{Aab}\psi_{ab},\qquad A=1,2,3.
\label{eq:conjugate-triplet}
\end{equation}
They couple to a remaining $\Z_3$ gauge field and form a triplet under the
new $\SU(3)_{\val}$ symmetry.  In this case translation is still realized
projectively, and pairing from this metal must break translation symmetry.
\end{itemize}

\textit{$\U(1)^2$ holon metal.---} If we remove the noncommuting terms in
the two adjoint Higgs fields above, we can reach a holon metal with two
$\U(1)$ gauge fields and three pockets related by translation symmetry.  In
the ordinary locking case $U=V$, the low-energy subspace is spanned by
$\psi_0,\psi_1,\psi_2$.  The gauge charges are most transparent in the basis
\begin{equation}
 \begin{pmatrix}\zeta_0\\ \zeta_1\\ \zeta_2\end{pmatrix}
 =\frac1{\sqrt3}
 \begin{pmatrix}
  1&1&1\\
  1&\omega&\omega^2\\
  1&\omega^2&\omega
 \end{pmatrix}
 \begin{pmatrix}\psi_0\\ \psi_1\\ \psi_2\end{pmatrix}.
\label{eq:charged-pocket-basis}
\end{equation}

The resulting low-energy theory is
\begin{align}
 \cL_{\rm h}={}&
 \sum_{r=0}^{2}\zeta_r^\dagger
 \left[\ii D_{t,r}+\frac{\bm D_r^2}{2m_*}+\mu_h\right]\zeta_r
 \nonumber\\
 &+\frac{1}{4\pi}K_{IJ}\alpha_I\dd\alpha_J
 +\frac{1}{12\pi}A\dd A,
\label{eq:holon-theory}\\
 D_{\mu,r}\zeta_r={}&
 (\partial_\mu-\ii Q_{rI}\alpha_{I\mu}+\ii A_\mu/3)\zeta_r .
\label{eq:holon-D}
\end{align}
Here $K=\left(\begin{smallmatrix}2&-1\\-1&2\end{smallmatrix}\right)$ is the
Cartan Chern--Simons matrix inherited from $SU(3)_1$ for the two $U(1)$
gauge fields $\alpha_1,\alpha_2$; repeated indices $I,J=1,2$ are summed.
The charge vectors are
$\bm Q_0=(1,0)^T$, $\bm Q_1=(-1,1)^T$, and $\bm Q_2=(0,-1)^T$.
Translations act as
$T_1\psi_j=\omega^j\psi_j$ and $T_2\psi_j=\omega^{-j}\psi_j$, and thus 
\begin{align}
T_1:\quad
 (\zeta_0,\zeta_1,\zeta_2)
 &\longmapsto
 (\zeta_1,\zeta_2,\zeta_0),\nonumber\\
 T_2:\quad
 (\zeta_0,\zeta_1,\zeta_2)
 &\longmapsto
 (\zeta_2,\zeta_0,\zeta_1).
\label{eq:holon-pocket-translations}
\end{align}

Let $\pi_T$ denote the permutation of charge sectors induced by a
translation, and let $\mathsf c:\bm Q_r\mapsto\bm Q_{r+1}$ be the cyclic
permutation.  Thus
 $(\pi_{T_1},\pi_{T_2})=(\mathsf c,\mathsf c^{-1})$.  Closely related
$U(1)^2$ theories appeared previously in Refs.~\cite{Zhang2025Holon,Fan2026},
with a one-sided nontrivial translation permutation, such as
$(\pi_{T_1},\pi_{T_2})=(1,\mathsf c)$ up to relabeling.  In the supplementary we
give a general proof that this realization is incompatible with $C_3$.
The construction here instead preserves the full triangular-lattice space
group.  The supplementary also shows
that the $U(1)^2$ holon metal must reduce $SU(3)_v$ to $S_3$.  It may
therefore be less natural for an ideal anyon gas, but can still be stabilized
on the triangular lattice when lattice-scale effects are appreciable.

\textit{Pairing instabilities of the $U(1)^2$ holon metal.---} Gauge fluctuations favor interpocket
pairing~\cite{Zhang2025Holon,Fan2026}.
The most singular current--current interaction contributes equally to the
$s$- and $p$-wave channels, so short-distance interactions are needed to
distinguish them.  For the present $C=+1/3$ quasihole orientation, the
density--current interaction generated by the Chern--Simons term favors
$p+ip$ pairing over $p-ip$.  Reversing the statistical angle reverses this
selection.  Within either channel,
there is close competition between a translation-invariant ansatz
\cite{Fan2026} and an ansatz that pairs only two pockets.  The latter yields
a charge-density-wave (CDW) metal~\cite{Zhang2025Holon}.  At leading BCS
order the two pairing patterns are degenerate, and nonlinear terms are needed to
select between them.  A CDW metal is therefore a strong competitor,
especially when the system already tends toward charge order.  Here we focus
on the translation-invariant state.  In the $p+ip$ channel, $\psi_1$ pairs
with $\psi_2$ while $\psi_0$ pairs within its own pocket.  The resulting
superconductor is the color-superconductor state with $N_\psi=3$ discussed
above.

In the $s$-wave channel, $\psi_1$ and $\psi_2$ pair, whereas $C_6$ leaves
the $\psi_0$ pocket unpaired.  The result is a gapless charge-$2e$
superconductor with a Bogoliubov Fermi surface.  Since
$c\sim\psi_0^\dagger\psi_1^\dagger\psi_2^\dagger$, after $\psi_1$ and
$\psi_2$ enter an $s$-wave paired state one may identify
$c\sim\psi_0^\dagger$.  Because $\psi_0$ remains unpaired,
$\langle cc\rangle=0$ even though the gauge-invariant charge-$2e$ composite
has condensed.  The coherent Bogoliubov Fermi surface therefore
produces a constant density of states at $E=0$ in tunneling measurements.

\textit{Chemical-potential-tuned transition out of the FCI.---} Equation~\eqref{eq:SU3-critical-theory} may describe a direct
chemical-potential-tuned continuous transition between the FCI and the
superconductor.  For $\mu<0$, the nine hole fields are gapped and the
remaining $\SU(3)_1$ theory describes the parent FCI.  For $\mu>0$, nine
small hole pockets appear and become unstable to gauge-mediated pairing,
producing a charge-$2e$ superconductor.  At $\mu=0$, the nine fermions have
quadratic dispersions with their band bottoms at zero energy.  This $z=2$
fixed point may be stable to pairing, a question we leave for future work.
If so, the FCI--SC transition can be continuous, with nine gapless fermions
at criticality.  For $\mu>0$, either $p+ip$ or $s$-wave pairing may be
favored; the latter gives a gapless superconductor with broken $SU(3)_v$.
One can alternatively consider an FCI--SC transition involving only two
$U(1)$ gauge fields and the three fermions $\psi_0,\psi_1,\psi_2$.  This
theory breaks the emergent $SU(3)_v$ while preserving triangular-lattice
symmetry.  It may be realizable on the triangular lattice but not on the
square lattice, allowing the two scenarios to be distinguished numerically.

This raises the question of whether all nine fermions must be retained in the
full theory.  Earlier composite-boson or composite-fermion approaches usually
produce only three pockets, whereas nine pockets have testable consequences.
The gauge-invariant particle--hole bilinear
${\cal M}_{bb'}=\sum_a\psi_{ab}^\dagger\psi_{ab'}$ describes density
fluctuations at momenta connecting distinct hole pockets.  If the paired gap
is small, an $SU(3)$ descendant exhibits low-energy density fluctuations at
$\mathbf Q=\Gamma,K,K'$ and at the six $\mathbf G/3$ momenta.  Descendants
of the $U(1)^2$ holon metal have such fluctuations only near
$\Gamma,K,K'$, while the one-pocket $Z_3$ metal has them only near
$\Gamma$.  The three-pocket $Z_3$ metal also has low-energy fluctuations at
$\mathbf Q=\mathbf G/3$, but its paired state must break translation
symmetry.  We therefore propose searching numerically and experimentally for
density fluctuations at $\mathbf G/3$, which can constrain the candidate
phases.

\textit{Conclusion.---} Doping the $C=1/3$ FCI supports a family of metallic
and superconducting phases.  A nine-fermion parton theory with
$SU(3)_g\times SU(3)_v$ provides a common framework for these states.
Color-superconducting pairing directly produces a charge-$2e$
superconductor with $c_-=1/2\pmod 1$.  Particle--hole Higgs terms instead
lead to $Z_3$ orthogonal metals or to the  $U(1)^2$
holon metal.  Our analysis shows that the $U(1)^2$ metal must break
$SU(3)_v$ and can only be stabilized on the triangular lattice when
lattice-scale effects are appreciable.  The nine-pocket $SU(3)$ theory may
also describe a chemical-potential-tuned FCI--superconductor transition.
Composite-boson and composite-fermion approaches can be interpreted as
selecting one or three pockets from the nine-pocket theory.  If all nine
fermions are required at the FCI--SC transition, the parton theory offers a
particularly natural framework.  It also enables future variational Monte
Carlo studies based on the unified wavefunction.

\textit{Acknowledgments.---} I thank T. Senthil, Ashvin Vishwanath, and
Taige Wang for inspiring discussions.
I used ChatGPT 5.6 Sol Extra High to assist with technical derivations.
The work is supported by the Alfred P. Sloan Foundation through a Sloan Research
Fellowship (Y.-H.Z.).

\bibliography{anyon_unification}

\clearpage
\onecolumngrid
\appendix
\setcounter{equation}{0}
\renewcommand{\theequation}{A\arabic{equation}}
\renewcommand{\theHequation}{A.\arabic{equation}}

\section*{Appendix A: From the triangular Hofstadter ansatz to the locked PSG}
\label{app:psg}

The projective matrices used in the main text are not independent symmetry
assumptions.  They follow from a microscopic parton hopping problem.  This
appendix derives them in three steps: first the real-space Hofstadter PSG of
$f$, then its action on the three hole valleys, and finally the new PSG after
the ordinary color--valley lock $U=V$.

\subsection*{A.1. Microscopic triangular-lattice Hofstadter ansatz}

Choose primitive vectors and the third nearest-neighbor direction
\begin{equation}
 \bm a_1=a(1,0),\qquad
 \bm a_2=a(1/2,\sqrt3/2),\qquad
 \bm a_3=\bm a_2-\bm a_1,\qquad
 \bm r_{xy}=x\bm a_1+y\bm a_2 .
\label{eq:app-hof-vectors}
\end{equation}
We take the oriented flux through a primitive rhombus to be
$\phi=-2\pi/3$; with our Berry-curvature convention this is the orientation
of the desired filled $C=1$ parton band.  Reversing $\phi$ complex conjugates
the PSG and reverses the band Chern numbers.  A convenient Landau gauge is
\begin{equation}
 A_1(x,y)=0,\qquad A_2(x,y)=\phi x,\qquad
 A_3(x,y)=\phi(x-2),
\label{eq:app-hof-gauge}
\end{equation}
where $A_j(\bm r)=A(\bm r,\bm r+\bm a_j)$.  Both elementary triangles have
Wilson loop $e^{-\ii\phi}$, while their product around the rhombus is
$e^{-2\ii\phi}=e^{\ii\phi}$ because $3\phi=-2\pi$.  Every Peierls phase is
therefore a power of the center element $\omega=e^{2\pi\ii/3}$.

For each color, the simplest Hofstadter Hamiltonian is
\begin{align}
 H_f={}&-t\sum_{a=1}^{3}\sum_{x,y}\Big[
 f_{a;x,y}^\dagger f_{a;x+1,y}
 +e^{\ii\phi x}f_{a;x,y}^\dagger f_{a;x,y+1}
 \nonumber\\
 &\hspace{31mm}
 +e^{\ii\phi(x-2)}f_{a;x,y}^\dagger f_{a;x-1,y+1}
 +\hc\Big]
 -\mu_f\sum_{a,x,y}f_{a;x,y}^\dagger f_{a;x,y}.
\label{eq:app-hof-H}
\end{align}
All hoppings are proportional to the color identity, so this saddle has
${\rm IGG}=\SU(3)_{\gauge}$.  Symmetry-preserving further-neighbor hopping
can isolate the desired filled $C=1$ band and arrange three low-energy
extrema without changing the PSG below.

The magnetic unit cell contains three sites.  Writing $x=3X+r$ with
$r=0,1,2$, define
\begin{equation}
 Q_\phi=\operatorname{diag}(1,e^{\ii\phi},e^{2\ii\phi}),\qquad
 \mathsf S(k_1)=
 \begin{pmatrix}0&1&0\\0&0&1\\e^{\ii k_1}&0&0\end{pmatrix}.
\label{eq:app-hof-magnetic-matrices}
\end{equation}
In this magnetic-sublattice basis the one-color Bloch Hamiltonian is
\begin{equation}
 h_f(\bm k)=-t\left[
 \mathsf S(k_1)+e^{\ii k_2}Q_\phi
 +e^{\ii k_2-2\ii\phi}Q_\phi\mathsf S^\dagger(k_1)
 +\hc\right]-\mu_f\mathbf1_3.
\label{eq:app-hof-Bloch}
\end{equation}
This is the microscopic band problem from which the continuum pocket fields
are projected.

\subsection*{A.2. Real-space magnetic PSG of the parton}

For a unitary lattice operation $g$, write its active action as
\begin{equation}
 \widetilde g f_{a;\bm r}\widetilde g^{-1}
 =z_g^{(f)}(\bm r)f_{a;g\bm r},\qquad
 z_g^{(f)}(\bm r)=e^{\ii\chi_g(\bm r)}\in\Z_3.
\label{eq:app-hof-active-PSG}
\end{equation}
The ansatz is invariant when
\begin{equation}
 A(g\bm r,g\bm r')-A(\bm r,\bm r')
 =\chi_g(\bm r')-\chi_g(\bm r)\pmod{2\pi}.
\label{eq:app-hof-link-condition}
\end{equation}
The coordinate generators are
\begin{align}
 T_1:(x,y)&\mapsto(x+1,y),&
 T_2:(x,y)&\mapsto(x,y+1),\nonumber\\
 C_6:(x,y)&\mapsto(-y,x+y),&
 M_x:(x,y)&\mapsto(x+y,-y).
\label{eq:app-hof-coordinate-actions}
\end{align}
Substitution into Eq.~\eqref{eq:app-hof-link-condition}, rather than an
abstract matrix guess, gives
\begin{align}
 \widetilde T_1 f_{a;x,y}\widetilde T_1^{-1}
 &=e^{\ii\phi y}f_{a;x+1,y},&
 \widetilde T_2 f_{a;x,y}\widetilde T_2^{-1}
 &=f_{a;x,y+1},
\label{eq:app-hof-translations}\\
 \widetilde C_6 f_{a;x,y}\widetilde C_6^{-1}
 &=e^{-\ii\phi[xy+(y^2+3y)/2]}
 f_{a;-y,x+y}.
\label{eq:app-hof-C6}
\end{align}
The exponent in Eq.~\eqref{eq:app-hof-C6} is always an integer.  Thus every
compensation is in the color center; no extra continuous $\U(1)$ gauge
factor has been introduced.

A unitary mirror reverses the flux.  The symmetry of one chiral Hofstadter
sector is instead the antiunitary magnetic mirror
$\widetilde M_x=M_x\mathcal K$.  Its link condition has a plus sign between
the two vector potentials, and one solution is
\begin{equation}
 \widetilde M_x f_{a;x,y}\widetilde M_x^{-1}
 =e^{\ii\phi(3y-y^2)/2}f_{a;x+y,-y},
 \qquad \widetilde M_x\ii\widetilde M_x^{-1}=-\ii.
\label{eq:app-hof-mirror}
\end{equation}
If the microscopic problem contains only a unitary $M_x$, this chiral saddle
breaks it, as expected for a state with nonzero Hall response.

For clarity, the four generator-dependent gauge factors in this Landau gauge
are
\begin{equation}
\boxed{
\begin{aligned}
 z_{T_1}^{(f)}(x,y)&=\omega^{-y},&
 z_{T_2}^{(f)}(x,y)&=1,\\
 z_{C_6}^{(f)}(x,y)&=
 \omega^{xy+(y^2+3y)/2},&
 z_{M}^{(f)}(x,y)&=\omega^{-(3y-y^2)/2}.
\end{aligned}}
\label{eq:app-hof-all-z}
\end{equation}
This also specifies $z_g^{(f)}$ for every group element, not only the
generators.  If $g_1$ is unitary,
\begin{equation}
 z_{g_1g_2}^{(f)}(\bm r)
 =z_{g_1}^{(f)}(g_2\bm r)z_{g_2}^{(f)}(\bm r),
\label{eq:app-hof-z-composition-unitary}
\end{equation}
whereas an antiunitary $g_1$ complex conjugates the coefficient produced by
$g_2$,
\begin{equation}
 z_{g_1g_2}^{(f)}(\bm r)
 =z_{g_1}^{(f)}(g_2\bm r)
 \left[z_{g_2}^{(f)}(\bm r)\right]^*.
\label{eq:app-hof-z-composition-antiunitary}
\end{equation}
The corresponding microscopic hole factor is
$z_g^{(\psi)}(\bm r)=[z_g^{(f)}(\bm r)]^*$.

The resulting operator algebra is
\begin{align}
 \widetilde T_1\widetilde T_2
 &=\omega^2\widetilde T_2\widetilde T_1,
\label{eq:app-hof-translation-algebra}\\
 \widetilde C_6\widetilde T_1\widetilde C_6^{-1}
 &=\widetilde T_2,&
 \widetilde C_6\widetilde T_2\widetilde C_6^{-1}
 &=\omega\widetilde T_1^{-1}\widetilde T_2,
\label{eq:app-hof-C6-algebra}\\
 \widetilde M_x\widetilde T_1\widetilde M_x^{-1}
 &=\widetilde T_1,&
 \widetilde M_x\widetilde T_2\widetilde M_x^{-1}
 &=\omega\widetilde T_1\widetilde T_2^{-1},
\label{eq:app-hof-mirror-algebra}\\
 \widetilde M_x^2&=1,&
 \widetilde M_x\widetilde C_6\widetilde M_x^{-1}
 &=\widetilde C_6^{-1}.
\label{eq:app-hof-dihedral-algebra}
\end{align}
In the real-space gauge above, $\widetilde C_6^6=1$ as well.
The center factors act nontrivially on one parton but disappear on the local
electron $c\sim f_1f_2f_3$ because $\omega^3=1$.  Equations
\eqref{eq:app-hof-translations}--\eqref{eq:app-hof-dihedral-algebra} are the
microscopic triangular-lattice PSG; all later matrices must represent this
algebra.

\subsection*{A.3. Projection to the three hole valleys}

Let \(u_b(\bm r)\), \(b=1,2,3\), be an orthonormal basis for the three
low-energy hole valleys of the Hofstadter band.  Define
\begin{equation}
 h_a[u]=\sum_{\bm r}u^*(\bm r)f_{a;\bm r},\qquad
 h_{ab}=h_a[u_b],\qquad \psi_{ab}=h_{ab}^\dagger .
\label{eq:app-projected-holes}
\end{equation}
For a unitary space-group operation \(g\), the microscopic PSG in A.2
induces
\begin{equation}
 (\mathsf d_gu)(\bm r)
 =\left[z_g^{(f)}(g^{-1}\bm r)\right]^*u(g^{-1}\bm r),
 \qquad
 (V_g^\dagger)_{cb}
 =\sum_{\bm r}u_c^*(\bm r)(\mathsf d_gu_b)(\bm r).
\label{eq:app-projected-V}
\end{equation}
For the antiunitary magnetic mirror, the last \(u\) in
Eq.~\eqref{eq:app-projected-V} is replaced by \(u^*\).  Since \(h_a[u]\) is
antilinear in \(u\), the hole matrix \(\Psi=(\psi_{ab})\) transforms as
\begin{equation}
 \widetilde g\,\Psi(\bm R)\widetilde g^{-1}
 =\Psi(g\bm R)V_g^\dagger .
\label{eq:app-unlocked-projected-action}
\end{equation}
Thus the position-dependent factor \(z_g^{(f)}(\bm r)\) is already included
in the constant valley matrix \(V_g\); it must not be applied again to the
slow field.

With \(\omega=e^{2\pi\ii/3}\), a convenient valley basis gives
\begin{align}
 V_{T_1}&=
 \begin{pmatrix}0&1&0\\0&0&1\\1&0&0\end{pmatrix},
 &V_{T_2}&=\operatorname{diag}(1,\omega,\omega^2),
 &V_{T_1}V_{T_2}&=\omega V_{T_2}V_{T_1},
\label{eq:app-valley-translations}\\
 V_{C_6}&=\frac{e^{-5\pi\ii/18}}{\sqrt3}
 \begin{pmatrix}
 \omega&\omega^2&\omega^2\\
 \omega&1&\omega\\
 \omega&\omega&1
 \end{pmatrix},
 &V_M&=\frac{e^{-\pi\ii/18}}{\sqrt3}
 \begin{pmatrix}
 1&\omega&\omega\\
 \omega&1&\omega\\
 \omega&\omega&1
 \end{pmatrix}.
\label{eq:app-valley-point-group}
\end{align}
The mirror acts antiunitarily.  These matrices are obtained by projecting
the real-space PSG of A.2 and satisfy the triangular magnetic-space-group
algebra up to the center of \(SU(3)\).

\subsection*{A.4. Locked PSG and its action on the nine fermions}

For the ordinary color--valley lock \(U=V\), a microscopic operation \(g\)
is accompanied by the color-gauge transformation \(U_g=V_g\).  Hence
\begin{align}
 \widetilde g_{\rm lock}:\quad
 \Psi(\bm R)&\longmapsto V_g\Psi(g\bm R)V_g^\dagger,
 &&g=T_1,T_2,C_6,\nonumber\\
 \widetilde M_{x,\rm lock}:\quad
 \Psi(\bm R)&\longmapsto
 V_M\Psi^*(M_x\bm R)V_M^\dagger .
\label{eq:app-locked-nine-action}
\end{align}
Although \(V_{T_1}\) and \(V_{T_2}\) commute only up to \(\omega\), the
center cancels between the left and right actions in
Eq.~\eqref{eq:app-locked-nine-action}.  The nine locked fermions therefore
carry ordinary crystal momenta.

In the basis \(\psi_0,\ldots,\psi_8\) defined in
Eq.~\eqref{eq:explicit-one-eight}, translations act diagonally:
\begin{align}
 T_1:\quad
 (\psi_0,\ldots,\psi_8)&\longmapsto
 (\psi_0,\omega\psi_1,\omega^2\psi_2,\psi_3,
 \omega\psi_4,\omega\psi_5,\psi_6,\omega^2\psi_7,\omega^2\psi_8),
 \nonumber\\
 T_2:\quad
 (\psi_0,\ldots,\psi_8)&\longmapsto
 (\psi_0,\omega^2\psi_1,\omega\psi_2,\omega^2\psi_3,
 \psi_4,\omega\psi_5,\omega\psi_6,\psi_7,\omega^2\psi_8).
\label{eq:app-nine-translations}
\end{align}
Suppressing the transformed spatial argument, the point group acts as
\begin{align}
 C_6:\quad&
 \psi_0\mapsto\psi_0,\qquad
 \psi_1\leftrightarrow\psi_2,\qquad
 \psi_3\to\psi_8\to\psi_7\to\psi_6\to\psi_5\to\psi_4\to\psi_3,
\label{eq:app-nine-C6}\\
 \widetilde M_x:\quad&
 \psi_0\mapsto\psi_0^*,\quad
 \psi_1\leftrightarrow\psi_7^*,\quad
 \psi_2\leftrightarrow\psi_4^*,\quad
 \psi_5\leftrightarrow\psi_8^*,\quad
 \psi_3\mapsto\psi_3^*,\quad
 \psi_6\mapsto\psi_6^* .
\label{eq:app-nine-mirror}
\end{align}
In Eq.~\eqref{eq:app-nine-mirror}, an expression such as
\(\psi_1\leftrightarrow\psi_7^*\) means
\(\psi_1\mapsto\psi_7^*\) and \(\psi_7\mapsto\psi_1^*\).  The star records
antiunitary conjugation of numerical coefficients; it does not turn a hole
annihilator into a creation operator.

\subsection*{A.5. The \texorpdfstring{$1+2+6$}{1+2+6} space-group split}

Equations~\eqref{eq:app-nine-translations}--\eqref{eq:app-nine-mirror}
organize the nine fermions into three closed space-group orbits,
\begin{equation}
 \{\psi_0\},\qquad
 \{\psi_1,\psi_2\},\qquad
 \{\psi_3,\psi_4,\psi_5,\psi_6,\psi_7,\psi_8\}.
\label{eq:app-one-two-six-orbits}
\end{equation}
Therefore the most general momentum-independent quadratic splitting that
preserves the full triangular magnetic space group is
\begin{equation}
 \delta H=
 \epsilon_0\,\psi_0^\dagger\psi_0
 +\epsilon_2\sum_{i=1}^{2}\psi_i^\dagger\psi_i
 +\epsilon_6\sum_{i=3}^{8}\psi_i^\dagger\psi_i .
\label{eq:app-one-two-six-splitting}
\end{equation}
After removing a common chemical-potential shift, this contains two
independent splittings and gives
\(\bm9\to\bm1\oplus\bm2\oplus\bm6\).
\setcounter{equation}{0}
\renewcommand{\theequation}{B\arabic{equation}}
\renewcommand{\theHequation}{B.\arabic{equation}}
\section*{Appendix B: Symmetry constraints on the
\texorpdfstring{$\U(1)^2$}{U(1)2} holon metal}
\label{app:u1-symmetry}

The three pockets of the \(\U(1)^2\) holon metal are
\(\zeta_0,\zeta_1,\zeta_2\), with gauge-charge vectors
\begin{equation}
 \bm Q_0=(1,0)^T,\qquad
 \bm Q_1=(-1,1)^T,\qquad
 \bm Q_2=(0,-1)^T .
\label{eq:app-U1-charges}
\end{equation}
A microscopic symmetry may permute these pockets while simultaneously
transforming the two \(\U(1)\) gauge fields.  We first show that any
nontrivial pocket permutation is incompatible with a continuous
\(SU(3)_v\) action and leaves at most \(S_3\).  We then determine which
permutations of \(T_1,T_2,C_6\) satisfy the triangular-lattice algebra.

\subsection*{B.1. Why a nontrivial translation permutation excludes
continuous \texorpdfstring{$SU(3)_v$}{SU(3)v}}

A symmetry acting on fields with \(\U(1)^2\) gauge charges must map every
allowed integer charge vector to another allowed integer charge vector.
Its action on gauge charge is therefore represented by
\begin{equation}
 M_g\in GL(2,\mathbb Z),\qquad
 \bm q\longmapsto M_g\bm q .
\label{eq:app-U1-charge-map}
\end{equation}
If the Chern--Simons matrix is \(K\), invariance further requires
\(M_gKM_g^T=K\).  The important point is that the allowed matrices \(M_g\)
form a discrete set.

Now consider any continuous valley rotation \(V(t)\), with
\(V(0)=I\), and let \(M(t)\) be its induced action on the \(\U(1)^2\)
gauge charges.  Because \(V(t)\) changes continuously while \(M(t)\) takes
values in a discrete set, \(M(t)\) cannot change along the path.  Therefore
\begin{equation}
 M(0)=I_2\quad\Longrightarrow\quad M(t)=I_2
 \quad\text{for the entire continuous path}.
\label{eq:app-U1-continuity}
\end{equation}
Every element of a continuous \(SU(3)_v\) is connected to the identity, so
a genuine \(SU(3)_v\) action must leave the \(\U(1)^2\) gauge charge
unchanged.  It can continuously rotate two pockets into one another only
when those pockets carry the same gauge charge.

Suppose instead that \(T_1\) nontrivially permutes the pockets, for example
\(\zeta_i\mapsto\zeta_j\) with \(i\ne j\).  Gauge covariance then requires
\begin{equation}
 M_{T_1}\bm Q_i=\bm Q_j .
\label{eq:app-U1-translation-charge-map}
\end{equation}
The three vectors in Eq.~\eqref{eq:app-U1-charges} are distinct, so
Eq.~\eqref{eq:app-U1-translation-charge-map} implies
\(M_{T_1}\ne I_2\).  If \(T_1\) could be embedded in a continuous
\(SU(3)_v\) rotation, a path from the identity to \(T_1\) would instead
force \(M_{T_1}=I_2\) by Eq.~\eqref{eq:app-U1-continuity}, which is a
contradiction.  The same argument applies when \(T_2\) acts by a nontrivial
permutation.

Thus a nontrivial permutation by either translation is incompatible with
continuous \(SU(3)_v\).  The remaining faithful action on the three pockets
is discrete and is at most their permutation group \(S_3\).  Appendix B.2
shows that translations realize its \(\Z_3\) subgroup and that \(C_6\)
enlarges it to the full \(S_3\).
\subsection*{B.2. Projective translations and sixfold rotation}

Let \(\pi_g\in S_3\) denote the permutation of
\((\bm Q_0,\bm Q_1,\bm Q_2)\) induced by a symmetry \(g\).  Let
\(\mathsf c\) be the three-cycle
\(\bm Q_r\mapsto\bm Q_{r+1}\), with \(r\) understood modulo three.
The triangular-lattice relations may be written
\begin{align}
 C_6T_1C_6^{-1}&=T_2,&
 C_6T_2C_6^{-1}&\doteq T_1^{-1}T_2 ,
\label{eq:app-U1-C6-algebra}\\
 C_3T_1C_3^{-1}&\doteq T_1^{-1}T_2,&
 C_3T_2C_3^{-1}&\doteq T_1^{-1},
\qquad C_3=C_6^2 .
\label{eq:app-U1-C3-algebra}
\end{align}
The symbol \(\doteq\) allows a \(\U(1)^2\) gauge transformation.  Such a
gauge transformation changes phases but does not permute the three charge
vectors, so it disappears from the permutation algebra.

Write \(\pi_i=\pi_{T_i}\) and \(\rho=\pi_{C_3}\).  Equation
\eqref{eq:app-U1-C3-algebra} gives
\begin{equation}
 \rho\pi_1\rho^{-1}=\pi_1^{-1}\pi_2,\qquad
 \rho\pi_2\rho^{-1}=\pi_1^{-1}.
\label{eq:app-U1-permutation-algebra}
\end{equation}
For a nontrivial cyclic realization, set
\(\pi_1=\mathsf c^{a_1}\), \(\pi_2=\mathsf c^{a_2}\), and
\(\rho\mathsf c\rho^{-1}=\mathsf c^\sigma\), where
\(\sigma=\pm1\).  Equation~\eqref{eq:app-U1-permutation-algebra} reduces to
\begin{equation}
 \sigma a_1=-a_1+a_2,\qquad
 \sigma a_2=-a_1\pmod3 .
\label{eq:app-U1-exponents}
\end{equation}
Its nontrivial \(C_3\)-symmetric solution is
\begin{equation}
 \boxed{(\pi_{T_1},\pi_{T_2})=(\mathsf c,\mathsf c^{-1})}
\label{eq:app-U1-inverse-translations}
\end{equation}
up to exchanging \(\mathsf c\leftrightarrow\mathsf c^{-1}\).
The alternatives \((1,\mathsf c)\) and \((\mathsf c,1)\) preserve
translations but violate Eq.~\eqref{eq:app-U1-permutation-algebra}; they are
therefore nematic.

Let \(s=\pi_{C_6}\).  Applying Eq.~\eqref{eq:app-U1-C6-algebra} to
Eq.~\eqref{eq:app-U1-inverse-translations} gives
\begin{equation}
 s\mathsf c s^{-1}=\mathsf c^{-1}.
\label{eq:app-U1-C6-inverts}
\end{equation}
Thus \(C_6\) acts as a transposition of two pockets.  Translations generate
the cyclic subgroup \(\langle\mathsf c\rangle\simeq\Z_3\), while including
\(C_6\) generates
\(\langle\mathsf c,s\rangle\simeq S_3\).  This is the projective
translation realization used in the main text.
\setcounter{equation}{0}
\renewcommand{\theequation}{C\arabic{equation}}
\renewcommand{\theHequation}{C.\arabic{equation}}
\section*{Appendix C: Pairing convention and Chern--Simons chirality selection}
\label{app:chirality}

Parameterize a counterclockwise momentum contour by
\begin{equation}
 \bm k=k(\cos\theta,\sin\theta),\qquad
 L_z=-\ii\partial_\theta .
\label{eq:app-Lz}
\end{equation}
We label a paired state by the winding of the anomalous expectation value
\begin{equation}
 F_\psi(\bm k)=\langle\psi_{\bm k}\psi_{-\bm k}\rangle
 \propto e^{+\ii\ell\theta}.
\label{eq:app-pair-winding}
\end{equation}
Thus \(\ell\) is the literal eigenvalue of
\(L_z=-\ii\partial_\theta\).  In this convention \(p+\ii p\) means
\(\ell=+1\) and \(F_\psi\propto k_x+\ii k_y\), while \(p-\ii p\) means
\(\ell=-1\) and \(F_\psi\propto k_x-\ii k_y\).

The corresponding mean-field Hamiltonian is the operator
\begin{align}
 H_{\rm BdG}={}&
 \sum_{\bm k}\xi_{\bm k}\psi_{\bm k}^\dagger\psi_{\bm k}
 +\frac12\sum_{\bm k}\left[
 \Delta_\ell(\bm k)\psi_{\bm k}^\dagger\psi_{-\bm k}^\dagger
 +\Delta_\ell^*(\bm k)\psi_{-\bm k}\psi_{\bm k}\right],
\label{eq:app-BdG-operator}\\
 \xi_{\bm k}={}&\frac{k^2}{2m}-\mu,\qquad
 \Delta_\ell(\bm k)\propto e^{+\ii\ell\theta}.
\nonumber
\end{align}
With the Nambu spinor
\(\Upsilon_{\bm k}=(\psi_{\bm k},\psi^\dagger_{-\bm k})^T\), its
single-particle BdG matrix is
\begin{equation}
 \mathcal H_\ell(\bm k)=
 \begin{pmatrix}
 \xi_{\bm k}&\Delta_\ell(\bm k)\\
 \Delta_\ell^*(\bm k)&-\xi_{\bm k}
 \end{pmatrix}.
\label{eq:app-BdG-matrix}
\end{equation}
This placement of the complex conjugates is important: in the BCS ground
state,
\begin{equation}
 F_\psi(\bm k)
 =\frac{\Delta_\ell(\bm k)}{2E_{\bm k}}
 \tanh\!\frac{E_{\bm k}}{2T},\qquad
 E_{\bm k}=\sqrt{\xi_{\bm k}^2+|\Delta_\ell(\bm k)|^2},
\label{eq:app-F-Delta}
\end{equation}
so \(F_\psi\) and \(\Delta_\ell\) have the same winding.

For completeness, write
\(\mathcal H_\ell=\bm d\cdot\bm\tau\), with
\(\bm d=(\operatorname{Re}\Delta_\ell,
-\operatorname{Im}\Delta_\ell,\xi)\).  If
\(\Delta_\ell=\Delta_0(k)e^{+\ii\ell\theta}\), then the azimuthal angle of
\(\hat{\bm d}\) winds as \(-\ell\theta\).  For \(\mu>0\),
\(\hat{\bm d}\) points south at \(k=0\) and north at \(k\to\infty\).
Using the filled-band convention
\(\mathcal A_i=-\ii\langle u_-|\partial_{k_i}u_-\rangle\) for the occupied,
negative-energy BdG eigenstate, one therefore finds
\begin{equation}
 C_{\rm BdG}
 =-\frac{1}{4\pi}\int\dd^2k\,
 \hat{\bm d}\cdot
 \left(\partial_{k_x}\hat{\bm d}\times
 \partial_{k_y}\hat{\bm d}\right)
 =-\ell .
\label{eq:app-Chern-derivation}
\end{equation}
Thus, in the convention of Eq.~\eqref{eq:app-pair-winding},
\begin{equation}
 \boxed{C_{\rm BdG}=-\ell,\qquad c_-^{\rm pocket}=-\frac{\ell}{2}.}
\label{eq:app-Cequalsl}
\end{equation}
For \(\mu<0\), the strong-pairing state has \(C_{\rm BdG}=0\).
Since the physical electron satisfies
\(c\sim\psi_1^\dagger\psi_2^\dagger\psi_3^\dagger\), a common hole winding
\(\ell\) gives the physical-electron angular momentum
\begin{equation}
 L_e=-3\ell .
\label{eq:app-electron-angular-momentum}
\end{equation}
In particular, hole \(p-\ii p\) has \(\ell=-1\) and
\(C_{\rm BdG}=+1\) per pocket, and
corresponds to physical-electron \(f+\ii f\).

We finally derive how the chirality is selected, starting from the
\(U(1)^2\) effective theory in Eqs.~\eqref{eq:holon-theory} and
\eqref{eq:holon-D}.  We work directly with its three fermions
\((\zeta_0,\zeta_1,\zeta_2)\).  The translation-preserving odd-parity ansatz
pairs the three inter-sector channels with equal amplitude,
\begin{equation}
 \langle\zeta_0(\bm k)\zeta_1(-\bm k)\rangle
 =\langle\zeta_1(\bm k)\zeta_2(-\bm k)\rangle
 =\langle\zeta_2(\bm k)\zeta_0(-\bm k)\rangle
 =\Delta e^{\ii\ell\theta}.
\label{eq:app-three-pocket-gap}
\end{equation}
Using the charge vectors in Eq.~\eqref{eq:app-U1-charges},
\begin{equation}
 K^{-1}=\frac13\begin{pmatrix}2&1\\1&2\end{pmatrix},\qquad
 \bm Q_r^T K^{-1}\bm Q_s=-\frac13\quad(r\ne s).
\label{eq:app-mutual-CS-charge}
\end{equation}

To fix the sign, start from the real-time action in
Eq.~\eqref{eq:holon-theory}, set the probe field to zero, and take
\(\epsilon_{012}=+1\).  Since
\(\ii D_{t,r}=\ii\partial_t+Q_{rI}\alpha_{I0}\), the density coupling is
\(+Q_{rI}\alpha_{I0}\rho_r\).  The corresponding kinetic Hamiltonian and
its term linear in the spatial gauge field are
\begin{equation}
 H_{{\rm kin},r}=\frac{(\bm p-Q_{rI}\bm\alpha_I)^2}{2m_*}-\mu_h,
 \qquad
 H_{{\rm int},r}=-\frac{Q_{rI}}{2m_*}
 \{\bm p,\bm\alpha_I\}.
\label{eq:app-real-time-current-vertex}
\end{equation}
Varying \(\alpha_{I0}\) gives
the Chern--Simons Gauss law
\begin{equation}
 \frac{K_{IJ}}{2\pi}b_J=-\sum_r Q_{rI}\rho_r.
\label{eq:app-CS-Gauss-law}
\end{equation}
In Coulomb gauge, a density in sector \(s\) therefore produces
\begin{equation}
 \alpha^{(s)}_{Ii}(\bm q)
 =-\frac{2\pi\ii}{q^2}(K^{-1})_{IJ}Q_{sJ}
 \epsilon_{ij}q_j\rho_s(\bm q),\qquad \epsilon_{xy}=+1.
\label{eq:app-CS-vector-potential}
\end{equation}
The term linear in \(\bm\alpha\) in the kinetic Hamiltonian gives two equal
density--current contributions to the Cooper process
\((\bm k,-\bm k)\to(\bm k',-\bm k')\).  Their sum is
\begin{equation}
 V_{\rm CS}^{rs}(\bm k',\bm k)
 =+\frac{4\pi\ii}{m_*}
 \left(\bm Q_r^TK^{-1}\bm Q_s\right)
 \frac{\bm k\mathbin{\times}\bm k'}{|\bm k'-\bm k|^2}.
\label{eq:app-CS-BCS-kernel-general}
\end{equation}
Here
\(\bm k\mathbin{\times}\bm k'\equiv
\epsilon_{ij}k_i k'_j=k_xk'_y-k_yk'_x\).  The sign follows directly from
the minus sign in Eq.~\eqref{eq:app-real-time-current-vertex} and the minus
sign in Eq.~\eqref{eq:app-CS-vector-potential}; the second
density--current contribution has the same sign as the first.
For the three inter-sector pairs in Eq.~\eqref{eq:app-three-pocket-gap}
and momenta on a circular Fermi surface, this becomes
\begin{equation}
 V_{\rm CS}(\vartheta)
 =-\ii\lambda_{\rm CS}\cot\frac{\vartheta}{2},
 \qquad \lambda_{\rm CS}=\frac{2\pi}{3m_*}>0,
\label{eq:app-CS-kernel}
\end{equation}
where \(\vartheta=\theta_{\bm k'}-\theta_{\bm k}\) is the outgoing angle
minus the incoming angle.  The coefficient can be renormalized by regular
gauge dynamics, but its sign is fixed by
Eq.~\eqref{eq:app-mutual-CS-charge}.  Written with the outgoing momentum
on the left, the linearized gap equation is
\begin{equation}
 \Delta_{rs}(\bm k')=-\int_{\bm k}
 V^{rs}(\bm k',\bm k)
 \frac{\tanh(\xi_{\bm k}/2T)}{2\xi_{\bm k}}\,
 \Delta_{rs}(\bm k),\qquad r\ne s.
\label{eq:app-linearized-gap-equation}
\end{equation}
Consequently an input gap \(e^{\ii\ell\theta_{\bm k}}\) is multiplied by
the partial wave
\(V_\ell=\operatorname{PV}\int_0^{2\pi}\frac{\dd\vartheta}{2\pi}
V_{\rm CS}(\vartheta)e^{-\ii\ell\vartheta}\).  Using
\begin{equation}
 \operatorname{PV}\int_0^{2\pi}\frac{\dd\vartheta}{2\pi}
 \cot\frac{\vartheta}{2}\,e^{-\ii\ell\vartheta}
 =-\ii\operatorname{sgn}(\ell),\qquad \ell\ne0,
\label{eq:app-cotangent-transform}
\end{equation}
one obtains
\begin{equation}
 V_\ell=-\lambda_{\rm CS}\operatorname{sgn}(\ell),
 \qquad \ell\ne0 .
\label{eq:app-CS-partial-waves}
\end{equation}
Indeed the linearized gap equation contains
\(-N_FV_\ell\ln(\Lambda/T)\), so a negative \(V_\ell\) is attractive.
The Chern--Simons term therefore selects
\(\ell=+1\), namely hole \(p+\ii p\), over \(p-\ii p\).  Reversing the
Chern--Simons level reverses this selection.  The \(s\)-wave channel is not
selected by Eq.~\eqref{eq:app-CS-kernel}; its competition with \(p\)-wave
pairing is controlled by the chirality-even gauge interaction and
short-distance interactions.  As a check, a dressed holon has exchange
angle
\(\theta_\psi=\pi-\pi\bm Q_r^TK^{-1}\bm Q_r
=\pi-2\pi/3=+\pi/3\pmod{2\pi}\), whose sign agrees with the selected
\(\ell=+1\) channel.
\setcounter{equation}{0}
\renewcommand{\theequation}{D\arabic{equation}}
\renewcommand{\theHequation}{D.\arabic{equation}}
\section*{Appendix D: Brief square-lattice comparison}
\label{app:square}

\subsection*{D.1. Explicit square-lattice realization}

For a square lattice with \(-2\pi/3\) flux per primitive cell, a convenient
Landau-gauge PSG is
\begin{align}
 \widetilde T_x f_{a;r,s}\widetilde T_x^{-1}
 &=\omega^{-s}f_{a;r+1,s},&
 \widetilde T_y f_{a;r,s}\widetilde T_y^{-1}
 &=f_{a;r,s+1},\nonumber\\
 \widetilde C_4 f_{a;r,s}\widetilde C_4^{-1}
 &=\omega^{rs}f_{a;-s,r},&
 \widetilde M_x f_{a;r,s}\widetilde M_x^{-1}
 &=f_{a;r,-s},
\label{eq:app-square-PSG}
\end{align}
where \(\widetilde M_x=M_x\mathcal K\) is antiunitary.  Projection to the
three valleys gives the same translation matrices as
Eq.~\eqref{eq:app-valley-translations}.  After the ordinary lock, every
unitary symmetry acts as
\(\Psi(\bm R)\mapsto V_g\Psi(g\bm R)V_g^\dagger\), so the center phase again
cancels.

Label the nine translation eigenfields by
\begin{equation}
 (\psi_0,\ldots,\psi_8)=
 (\chi_{00},\chi_{11},\chi_{22},\chi_{10},\chi_{01},
 \chi_{21},\chi_{20},\chi_{02},\chi_{12}),
\label{eq:app-square-field-map}
\end{equation}
where
\begin{equation}
 T_x:\chi_{mn}\mapsto\omega^n\chi_{mn},\qquad
 T_y:\chi_{mn}\mapsto\omega^{-m}\chi_{mn}.
\label{eq:app-square-translations}
\end{equation}
If \(\bm b_\mu\cdot\bm a_\nu=2\pi\delta_{\mu\nu}\), the corresponding
momenta are
\begin{equation}
 \bm k_{mn}=\frac{n\bm b_x-m\bm b_y}{3}
 \quad\text{modulo reciprocal lattice vectors}.
\label{eq:app-square-momenta}
\end{equation}
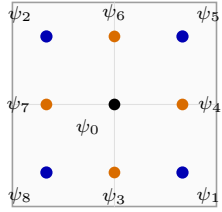
\begin{figure}[h]
\centering
\begin{tikzpicture}[
  x=0.90cm,y=0.90cm,
  every node/.style={font=\scriptsize}
]
  \fill[black!2] (-1.5,-1.5) rectangle (1.5,1.5);
  \draw[black!38,semithick] (-1.5,-1.5) rectangle (1.5,1.5);
  \draw[black!12,thin] (-1.5,0)--(1.5,0);
  \draw[black!12,thin] (0,-1.5)--(0,1.5);

  \fill (0,0) circle (2.2pt);
  \node[anchor=north east] at (-0.08,-0.08) {$\psi_0$};

  \fill[orange!85!black] (1,0) circle (2.2pt);
  \node[anchor=west] at (1.10,0) {$\psi_4$};
  \fill[orange!85!black] (-1,0) circle (2.2pt);
  \node[anchor=east] at (-1.10,0) {$\psi_7$};
  \fill[orange!85!black] (0,1) circle (2.2pt);
  \node[anchor=south] at (0,1.10) {$\psi_6$};
  \fill[orange!85!black] (0,-1) circle (2.2pt);
  \node[anchor=north] at (0,-1.10) {$\psi_3$};

  \fill[blue!70!black] (1,-1) circle (2.3pt);
  \node[anchor=north west] at (1.08,-1.08) {$\psi_1$};
  \fill[blue!70!black] (-1,1) circle (2.3pt);
  \node[anchor=south east] at (-1.08,1.08) {$\psi_2$};
  \fill[blue!70!black] (1,1) circle (2.3pt);
  \node[anchor=south west] at (1.08,1.08) {$\psi_5$};
  \fill[blue!70!black] (-1,-1) circle (2.3pt);
  \node[anchor=north east] at (-1.08,-1.08) {$\psi_8$};
\end{tikzpicture}
\caption{\label{fig:square-nine-field-momenta}
The nine square-lattice momenta.  The center is $\psi_0$; orange (blue)
points form the four-field axis (diagonal) orbit.}
\end{figure}
The square point group acts on the labels as
\begin{equation}
 C_4:(m,n)\mapsto(n,-m),\qquad
 \widetilde M_x:(m,n)\mapsto(m,-n),
\label{eq:app-square-label-action}
\end{equation}
with the mirror also conjugating numerical coefficients.  Consequently the
nine fields form the three space-group orbits
\begin{equation}
 \{\psi_0\},\qquad
 \{\psi_3,\psi_4,\psi_6,\psi_7\},\qquad
 \{\psi_1,\psi_2,\psi_5,\psi_8\}.
\label{eq:app-square-orbits}
\end{equation}
Thus the square-lattice splitting is
\(\bm9\to\bm1\oplus\bm4_{\rm axis}\oplus\bm4_{\rm diagonal}\).
In particular, no three-field subset is closed under the full square space
group, so the triangular-lattice three-pocket \(\U(1)^2\) holon metal has no
fully square-symmetric counterpart.

\subsection*{D.2. General obstruction for symmetric $U(1)^2$ holon metal}

  Let \(p_x,p_y,s\in S_3\) be the
permutations of the three distinct \(\U(1)^2\) charge sectors induced by
\(T_x,T_y,C_4\).  Gauge transformations can change phases, but not these
permutations.  Because the translation commutator lies in the IGG,
\(p_xp_y=p_yp_x\).  The rotation algebra gives
\begin{equation}
 s p_xs^{-1}=p_y,\qquad
 s p_ys^{-1}=p_x^{-1}
 \quad\Longrightarrow\quad
 s^2p_xs^{-2}=p_x^{-1}.
\label{eq:app-square-permutation-algebra}
\end{equation}
For a translation-related three-pocket metal, take
\(p_x=\mathsf c^a\), where \(\mathsf c\) is a three-cycle and \(a=1,2\).
For every \(s\in S_3\), \(s^2\in A_3=\{1,\mathsf c,\mathsf c^2\}\), so
\(s^2\) commutes with \(p_x\).  Equation~\eqref{eq:app-square-permutation-algebra}
would then require \(p_x=p_x^{-1}\), impossible for a three-cycle.  Hence a
cyclic three-pocket translation realization is incompatible with \(C_4\).
The other solutions have trivial translation permutations or the same
transposition for both translations, and therefore do not put all three
charge sectors in one translation orbit.
\end{document}